\documentclass[12pt]{iopart}
\usepackage{iopams}  

\usepackage{graphicx} 
\usepackage{dcolumn} 
\usepackage{bm} 
\usepackage{color} 

\usepackage{hyperref}
\usepackage{doi}

\expandafter\let\csname equation*\endcsname\relax

\expandafter\let\csname endequation*\endcsname\relax

\usepackage{amsmath}

\usepackage{soul}

\usepackage{amssymb}
\usepackage{mathrsfs}
\usepackage{upgreek}
\usepackage{marvosym}
\usepackage{siunitx}
\usepackage{multirow}

\definecolor{darkpastelgreen}{rgb}{0.01, 0.75, 0.24}

\begin{document}

\title[CDW proximity effects on SO and exchange coupling in Gr on 1T-TaS$_2$ monolayer]{Spin-orbit and exchange proximity couplings in graphene/1T-TaS$_2$ heterostructure triggered by a charge density wave}

\author{Karol Sza\l{}owski$^1$, Marko Milivojevi{\'c}$^2$, Denis Kochan$^3$, and Martin Gmitra$^4$}

    \address{$^1$ University of \L{}\'od\'z, Faculty of Physics and Applied Informatics, Department of Solid State Physics, PL90-236 \L{}\'od\'z, Poland}
        \ead{karol.szalowski@uni.lodz.pl}

\address{$^2$ Institute of Informatics, Slovak Academy of Sciences, 84507 Bratislava, Slovakia\\
Institute for Theoretical Physics, University of Regensburg, 93053 Regensburg, Germany\\
        Faculty of Physics, University of Belgrade, 11001 Belgrade, Serbia}
	    \ead{milivojevic@rcub.bg.ac.rs}

	\address{$^3$ Institute of Physics, Slovak Academy of Sciences, 84511 Bratislava, Slovakia\\
    Institute for Theoretical Physics, University of Regensburg, 93053 Regensburg, Germany}
	    \ead{denis.kochan@savba.sk}
	
	\address{$^4$ Institute of Physics, Pavol Jozef \v{S}af\'{a}rik University in Ko\v{s}ice, 04001 Ko\v{s}ice, Slovakia\\
	Institute of Experimental Physics, Slovak Academy of Sciences, 04001 Ko\v{s}ice, Slovakia}
	\ead{martin.gmitra@upjs.sk}

\vspace{10pt}
\begin{indented}
\item[]\today
\end{indented}

\begin{abstract}
Proximity-induced fine features and spin-textures of the electronic bands in graphene-based van der Waals heterostructures can be explored from the point of tailoring a twist angle. Here we study spin-orbit coupling and exchange coupling engineering of graphene states in the proximity of 1T-TaS$_2$ not triggering the twist, but a charge density wave in 1T-TaS$_2$---a realistic low-temperature phase. Using density functional theory and effective model we found that the emergence of the charge density wave in 1T-TaS$_2$ significantly enhances Rashba spin-orbit splitting in graphene and tilts 
the spin texture by a significant Rashba angle---in a very similar way as in the conventional twist-angle scenarios. 
Moreover, the partially filled Ta $d$-band in the charge density wave phase leads to the spontaneous emergence of the in-plane magnetic order that transgresses via proximity from 1T-TaS$_2$ to graphene, hence, simultaneously superimposing along the spin-orbit 
also the exchange coupling proximity effect. To describe this intricate proximity landscape we have developed an effective model Hamiltonian and provided a minimal set of parameters that excellently reproduces all the spectral features predicted by the first-principles calculations.
Conceptually, the charge density wave provides a highly interesting knob to control the fine features of electronic 
states and to tailor the superimposed proximity effects---a sort of \emph{twistronics without twist}. 
\end{abstract}

\section{Introduction}

The ability to stack two-dimensional materials offers to design ultrathin van der Waals heterostructures \cite{Geim2013,Sierra2021,Kurebayashi2022} that can foster spintronics \cite{Han2014,Avsar2014,Han2015,Wang2021,Galceran2021,Gmitra2017}, optoelectronics \cite{Gmitra2015,Gmitra2017,Liao2019}, and twistronics \cite{Carr2017,Hennighausen2021} applications and pave the way to novel carbon-based and, hence, environmentally-friendly devices.
Modifying the properties of van der Waals materials through proximity effects is an added degree of functionality so far 
very intensively explored, particularly, proximitizing graphene leads to the appearance of induced spin-orbit coupling (SOC) \cite{Gmitra2015}, or inducing magnetic exchange proximity effects \cite{Zollner2016,Zollner2020}.

In order to add a novel tuning knob to the van der Waals heterostructures, we investigate the effect of charge density wave (CDW) that develops in 1T-TaS$_2$ and explore its effect on the spin degree of freedom in proximitized graphene. 
In addition to various CDW orderings that are emerging in layered 3D bulk structures of transition metal dichalcogenides (TMDC) \cite{Rossnagel2011}, also their 
2D counterparts \cite{Lin2020NP,Xu2021,Lasek2021} offer a potential platform for studying the effects stemming from the dimensionality, and CDW-triggered commensurability and 
shape-topography \cite{Ngankeu2017,Martino2020}.
In particular, some of TMDC in the 1T polymorph phase have the propensity to exhibit a variety of periodic lattice distortions \cite{Pasquier2019}. Frequently, the Peierls instability gives rise to a stable low temperature periodic CDW phase with
a triangular $\sqrt{13}\times\sqrt{13}$ superlattice. Particular examples count 1T-TMDC such TaS$_2$ \cite{Wilson1975,Brouwer1980,Miller2018}, TaSe$_2$ \cite{Miller2018,Zhang2020,Jiang2021} and TaTe$_2$ \cite{Miller2018,Jiang2021} or NbS$_2$ \cite{Tresca2019} and NbSe$_2$ \cite{Calandra2018,Liu2021,Liu2021b}. Structural CDW distortions have the form of David stars \cite{Sipos2008}---clusters of 13  transition metal atoms in which an excess electron is effectively trapped at the star center \cite{Tosatti1976,Fazekas1979,Fazekas1980,Smith1985,Rossnagel2006,Kratochvilova2017}. What is particularly interesting, the commensurate $\sqrt{13}\times\sqrt{13}$ CDW phase of TaS$_2$ forming the superlattice of David stars \cite{Ritschel2013} possesses below 180~K \cite{Perfetti2005} enough electronic correlations \cite{Kim2022} to develop the Mott phase \cite{Shin2021,Petkov2022,Petocchi2022,Ren2022,Fei2022} and even superconductivity under pressure \cite{Sipos2008,Darancet2014}. 
Formation of CDW in 1T-TaS$_2$ results from the subtle interplay of lattice, orbital and spin degrees of freedom \cite{Yi2018}, along with the electron-phonon coupling \cite{Gruner1988,Liu2009} which results in strongly localized electronic states near the Fermi level \cite{Wen2021}. The anharmonic phonon-phonon interactions in 1T-TaS$_2$ are predicted to stabilize CDW at elevated temperatures \cite{Lazar2015}.
The correlations at low-temperatures lead to spin polarization and transform a monolayer of 1T-TaS$_2$ into a magnetically polarized semiconductor~\cite{Zhang2014} with in-plane magnetization \cite{Pasquier2022}. 
Rich electronic properties of 1T-TaS$_2$ can be engineered by the temperature~\cite{Lutsyk2018,Wang2020,Mohammadzadeh2021,Mihailovic2021}, pressure~\cite{Lee2019}, electrical pulses~\cite{Hollander2015,Ma2016}, electrical bias~\cite{Zhao2018,Li2021}, creation of domain electronic walls~\cite{Cho2017}, insertion of adsorbates~\cite{Lee2020,Lee2022a}, charge doping~\cite{Shao2016}, substrate selection~\cite{Zhao2017}, strain~\cite{Gan2016,Bu2019,Nicholson2022}, or laser light~\cite{Stojchevska2014,Ludwiczak2020,Lacinska2022,Ren2023}. All this gives the 1T-TaS$_2$ monolayer a potential for 2D spintronics applications counting ionic field-effect transistor~\cite{Yu2015}, logic gates and circuits~\cite{Khitun2018},
non-volatile memory devices~\cite{Vaskivskyi2015,Vaskivskyi2016,Mihailovic2021},
and others.

In addition to many promising studies of 1T-TaS$_2$ in the bulk or monolayer form, also its various heterostructures have been considered in recent literature, including the experimental works on the heterostructures with 2H-TaS$_2$ \cite{Ayani2022}, 2H-MoSe$_2$ and 2H-WSe$_2$ \cite{Lee2022}, 2H-MoS$_2$ \cite{Grisafe2018}, permalloy \cite{Husain2021}, h-BN \cite{Taheri2022} or black phosphorus \cite{Wang2019}, and also graphene \cite{Altvater2021,Altvater2022,Kim2022,Boix-Constant2021}. The latter is providing an ideal platform for investigating fine spectral features stemming from the proximity of graphene Dirac electrons to the correlated CDW phase. 

The CDW degree of freedom present in 1T-TaS$_2$ and its spatial topology provides a channel to control the proximity exchange coupling and spin-orbit coupling in graphene without any physical change in a twist of the heterostructure components, as the CDW phase can be manipulated locally, for example, by optical means~\cite{Zhu2018,Laulhe2017,Li2020}, electrical pulses~\cite{Vaskivskyi2015,Vaskivskyi2016}, or just temperature and in a fully reversible manner. Such a feature has been only achieved in optical lattice analog to bilayer graphene~\cite{Salamon2020}. Twisted graphene/TMDC heterostructures have been recently intensively studied~\cite{David2019,Li2019b,Pezo2021,Zollner2021,Naimer2021} and control of the twist angle is of paramount importance for possible applications based on the design of spin-charge conversion utilizing the Rashba-Edelstein effect~\cite{Galceran2021,Ingla-Aynes2022,Li2020,Veneri2022,Lee2022b}. Here we show that the same level of functionality can be expected also for graphene/CDW states where
the ``role of the twist'' is played by the spatial topology of CDW.

In the paper, we study the van der Waals heterostructure of graphene and 1T-TaS$_2$ monolayer by means of first-principles calculations and propose an effective tight-binding model for graphene bands near the Dirac point extracting relevant model parameters. 
We found that the model parameters reflect induced spin-orbit coupling and exchange interaction proximity effects for transitions of the 1T-TaS$_2$ monolayer from the normal to CDW state, and then to the ferromagnetic ground state achieved by gradually decreasing temperature.

\section{Effective Model Hamiltonian}
Based on a given structural symmetry one can derive a specific effective model Hamiltonian for graphene $\pi$-bands \cite{Kochan2012,Kochan2017}. It has been noted \cite{David2019,Li2019b} that the most general form of the induced Rashba spin-orbit coupling term in twisted graphene/TMDC heterostructures that obeys time reversal and threefold rotation $C_3$ symmetries can be written as $U^\dagger H_{\rm R} U$, where $H_{\rm R}$ is the effective Rashba Hamiltonian, $H_{\rm R} = \lambda_{\rm R}(\kappa \sigma_x s_y - \sigma_y s_x)$ \cite{Kane2005,Bychkov1984} and $U=e^{-i s_z \phi_{\rm R}/2}$, where $\sigma_{x,y}$ are the Pauli matrices acting in sublattice space, while $s_{x,y,z}$ are the Pauli spin matrices for the spin degree of freedom. The Rashba angle $\phi_{\rm R}$ originates from spin-orbit coupling matrix elements between $d$-orbitals of metal atoms in TMDC \cite{David2019} and in general, represents a sum of the geometric angle and the quantum phase \cite{Peterfalvi2022}.
The results provided below are given with respect to the ordered 
Bloch basis: 
\begin{equation}\label{basis}
|A(\mathbf{q}),\uparrow\rangle, |A(\mathbf{q}),\downarrow\rangle, |B(\mathbf{q}),\uparrow\rangle, |B(\mathbf{q}),\downarrow\rangle,
\end{equation}
where the states are sublattice (A, B) and spin ($\uparrow$, $\downarrow$) resolved, and the crystal momentum $\mathbf{q}$ is measured from the $\Gamma$ point.

The corresponding effective Hamiltonian in the studied cases has $C_3$ symmetry, and its low energy expansion close to the K $(\kappa=1)$ and K' $(\kappa=-1)$ point is given as follows:
\begin{eqnarray}
\label{Eq:Ham}
\fl H_{\kappa}=
\left(
\begin{matrix}
E_0+\Delta+\kappa \lambda_{\rm I}^{\rm A} & 0 & \frac{\sqrt{3}}{2}a t \left(\kappa k_x-ik_y\right) & i(1-\kappa)\lambda_{\rm R} e^{-i\phi_{\rm R}} \\
0 & E_0+\Delta-\kappa \lambda_{\rm I}^{\rm A} & i(1+\kappa)\lambda_{\rm R} e^{i\phi_{\rm R}} & \frac{\sqrt{3}}{2}a t \left(\kappa k_x-ik_y\right) \\
\frac{\sqrt{3}}{2}a t \left(\kappa k_x+ik_y\right) & -i(1+\kappa)\lambda_{\rm R} e^{-i\phi_{\rm R}} &
E_0-\Delta-\kappa \lambda_{\rm I}^{\rm B} & 0\\
-i(1-\kappa)\lambda_R e^{i\phi_{\rm R}} & \frac{\sqrt{3}}{2}a t \left(\kappa k_x+ik_y\right) &
0 & E_0-\Delta+\kappa \lambda_{\rm I}^{\rm B}
\end{matrix}\right).
\end{eqnarray}
The orbital and spin proximity effects result in the following parameters: $E_0$ corresponds to the Dirac point offset with respect to the Fermi energy, $\Delta$ is the staggered potential discriminating electronic states located at sublattice A and B, $\lambda_{\rm I}^{\rm A}$ and $\lambda_{\rm I}^{\rm B}$ are the sublattice-resolved spin-conserving next-nearest neighbor spin-orbit couplings, $\lambda_{\rm R}$ is the Rashba spin-orbit coupling parameter with phase angle $\phi_{\rm R}$. Moreover, orbital nearest neighbor hopping $t$ and the lattice constant $a$ complete the set of parameters for the Hamiltonian as those can be subjected to CDW distortion. We should mention that the sublattice-resolved SOC parameters can be expressed as $\lambda_{\rm I}^{\rm A}=\lambda_{\rm I}+\delta\lambda_{\rm I}$ and $\lambda_{\rm I}^{\rm B}=\lambda_{\rm I}-\delta\lambda_{\rm I}$, where $\lambda_{\rm I}$ is conventionally called intrinsic, or Kane-Mele SOC.

The above-described Hamiltonian (\ref{Eq:Ham}) was used to model graphene band structure in the vicinity of the K/K' points for normal and CDW phase of 1T-TaS$_2$. The model parameters were determined to fit the density functional theory (DFT) calculations. A partially filled band in the vicinity of the Fermi level tends to spin split \cite{Zhang2014} at low temperatures and strong correlations of the Ta $d$-electrons play an important role in the positioning of the lower and upper Hubbard bands \cite{Ang2012,Darancet2014,Wang2020,Butler2020,Altvater2022}. This has an effect on the emergence of the in-plane magnetic polarization of the 1T-TaS$_2$~\cite{Pasquier2022}. In order to describe the spin-orbit coupling and also
the exchange proximity coupling of the graphene bands around the Dirac points---we model the magnetic interaction between graphene and 1T-TaS$_2$ Hubbard band in terms of effective Zeeman-like interactions captured by the following Hamiltonian:
\begin{equation}\label{Eq:Ham_ex}
H_{\rm ex}=\begin{pmatrix}
0&\Delta_{\rm A}{\rm e}^{-{\rm i}\phi_{\rm A}}&0&0\\
\Delta_{\rm A}{\rm e}^{{\rm i}\phi_{\rm A}} & 0 & 0 & 0 \\
0 & 0 & 0 & \Delta_{\rm B}{\rm e}^{-{\rm i}\phi_{\rm B}} \\
0 & 0 & \Delta_{\rm B}{\rm e}^{{\rm i}\phi_{\rm B}} & 0
\end{pmatrix}.
\end{equation}
The effective Hamiltonian $H_{\rm ex}$, independent of the valley, represents the coupling of a spin with the sublattice-resolved (subscripts A/B) in-plane magnetization that is parameterized by the amplitude $\Delta_{\rm A/B}$ and the angle $\phi_{\rm A/B}$ 
between the $x$-axis and the magnetization direction.

\section{First-principles calculations details}
\begin{figure}
    \centering
    \includegraphics[width=0.75\columnwidth]{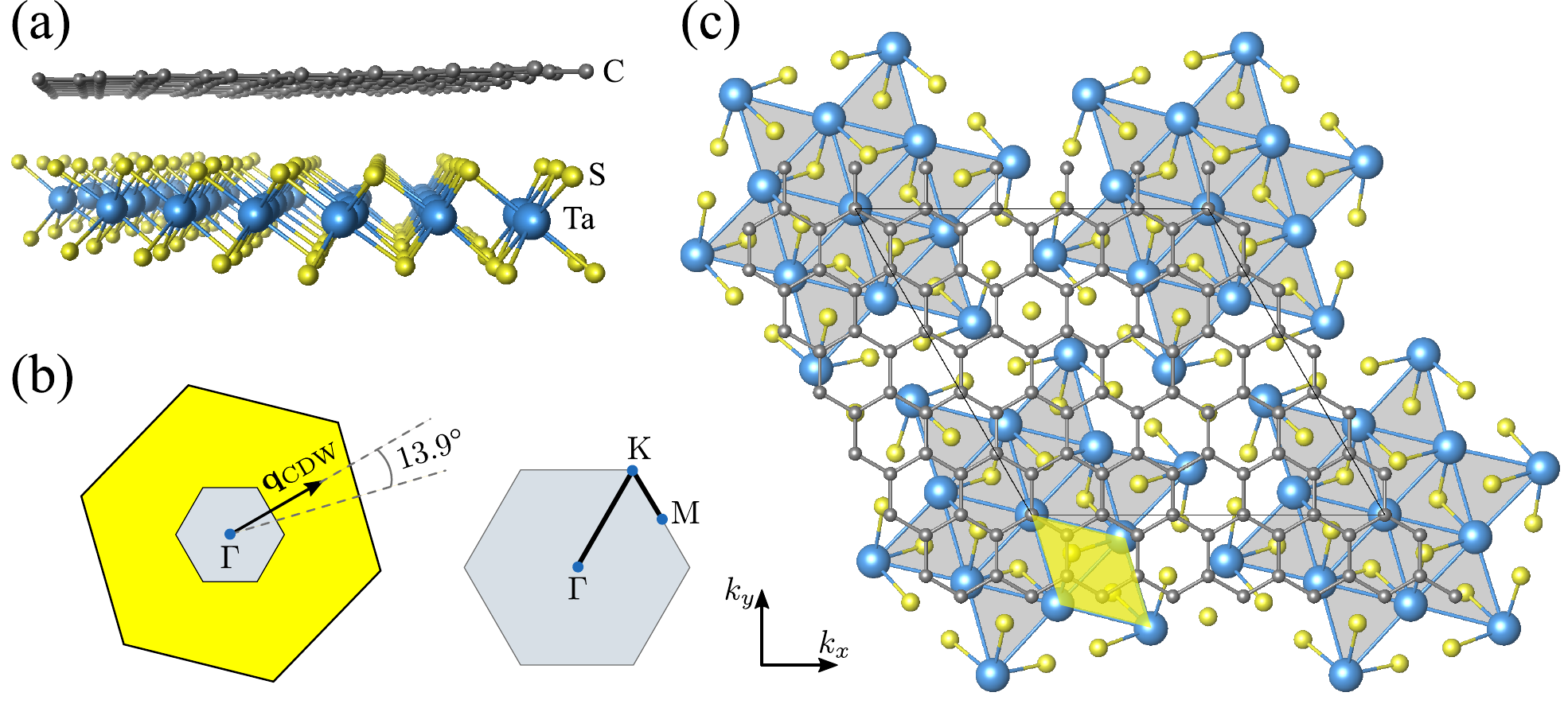}
    \caption{Structure details of graphene/1T-TaS$_2$ heterostructure. (a)~side view of the heterostructure. (b)~first Brillouin zone of considered heterostructure (gray) with charge density wave vector $\mathbf{q}_{\rm CDW}$ and the Brillouin zone of the undistorted  1T-TaS$_2$ (yellow). Sketch of the graphene Brillouin zone with the high-symmetry path used in band structure calculations. (c)~top view of the supercell with charge density wave and graphene in top stacking with carbon atom centered in the middle of a David star (grayish background).}
    \label{fig1}
\end{figure}

We consider a heterostructure with a structural model formed by the supercell of $\sqrt{13}\times\sqrt{13}$ for 1T-TaS$_2$ and $5\times 5$ for graphene, see Fig.~\ref{fig1}. We study a slab geometry with a vacuum of about 15~${\rm\AA}$ in the direction perpendicular to the planes in order to separate periodic images. The graphene lattice parameter was compressed by approximately 1.4\% to appropriately fit the supercell, capturing the commensurate CDW in 1T-TaS$_2$, using the experimental lattice constant of 12.1323 {\AA} \cite{Spijkerman1997}. 
The commensurate CDW can be described by a wavector $\mathbf{q}_{\rm CDW}$ whose orientation stems from the emerging of $\sqrt{13}\times\sqrt{13}$ reconstruction and rotation angle of 13.9$^{\circ}$ \cite{Spijkerman1997,Stahl2020}, see Fig.~\ref{fig1}(b).

All the electronic structure calculations were performed using the {\sc Quantum Espresso} suite~\cite{Giannozzi2009,Giannozzi2017}, implementing a plane wave basis for DFT calculations~\cite{Hohenberg1964}. The Perdew-Burke-Ernzerhof exchange-correlation functional was utilized~\cite{Perdew1996} for the projector augmented wave method~\cite{Kresse1999}. For non-collinear DFT calculations including spin-orbit coupling, fully relativistic pseudopotentials were applied~\cite{DalCorso2014}. Small Methfessel-Paxton energy level smearing~\cite{Methfessel1989} of 1~mRy was used along with the kinetic energy cut-off of 53~Ry. The semiempirical van der Waals corrections were included~\cite{Grimme2006,Barone2009} and the dipole correction~\cite{Bengtsson1999} was applied to properly determine Dirac point energy offset due to dipole electric field effects. The relaxation of atomic positions within the spin-unpolarized DFT approach was allowed with a threshold force of $10^{-4}$ Ry per Bohr radius. For spin-polarized DFT+U calculations~\cite{Cococcioni2005}, we used on-site Hubbard $U = 3$~eV for Ta $d$-orbitals, inspired by the calculations reported in Refs.~\cite{Martino2020,Boix-Constant2021}, with $9\times 9$ mesh of $k$-points used for sampling the first Brillouin zone. 

\section{Electronic band structures and results of fitting}
We study two stacking configurations of graphene on 1T-TaS$_2$, particularly, the \emph{top position} where the carbon atom is placed over the Ta atom of the David star center and the \emph{hollow position} where the center of graphene ring and the David star are placed on top of each other, see insets in Fig.~\ref{fig3}.
The CDW formation lowers the heterostructure energy per supercell by 152 meV for the top and by 136 meV for the hollow stacking, respectively. We note that those energies reasonably correspond to the transition of 1T-TaS$_2$ to the commensurate CDW phase \cite{Sipos2008}.

\begin{figure}
    \centering
    \includegraphics[width=1.0\columnwidth]{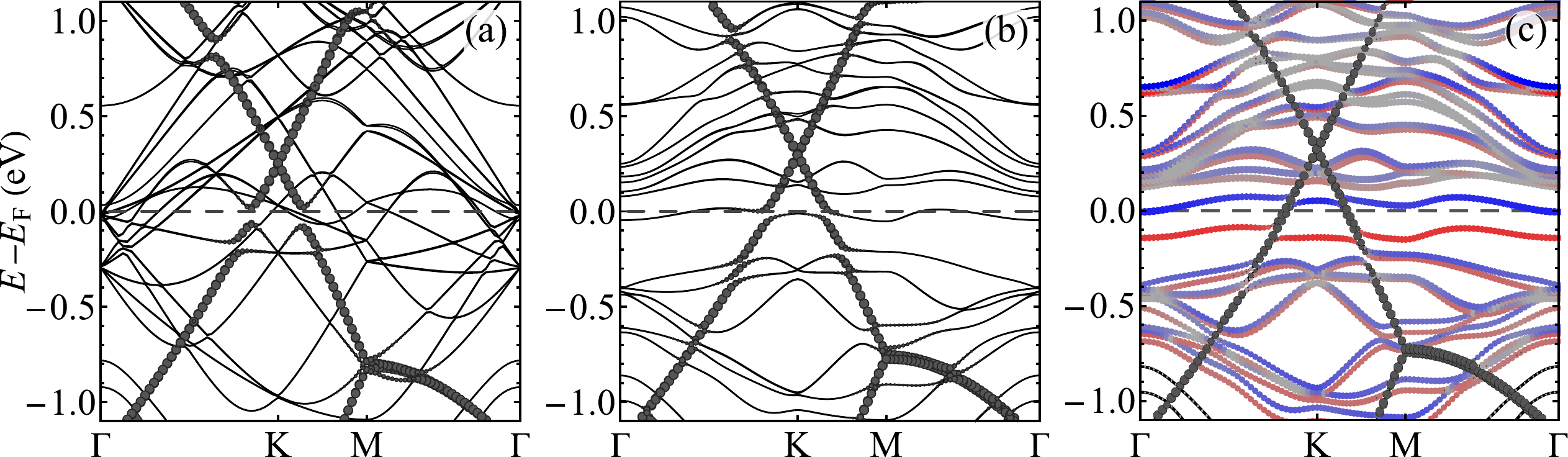}
    \caption{Calculated electronic band structure of graphene/1T-TaS$_2$ heterostructure in the top configuration for normal state (a), CDW state (b) and CDW magnetic state (c) of the 1T-TaS$_2$ monolayer. The fat bands for graphene states are shown with filled dark gray circles. The fat bands from Ta atoms are shown in (c) with filled circles, where the color indicates the degree of spin polarization (ranging within a linear scale from red for fully polarized spin-up through gray to blue for spin-down states).}
    \label{fig2}
\end{figure}

\begin{figure}
    \centering
    \includegraphics[width=0.99\columnwidth]{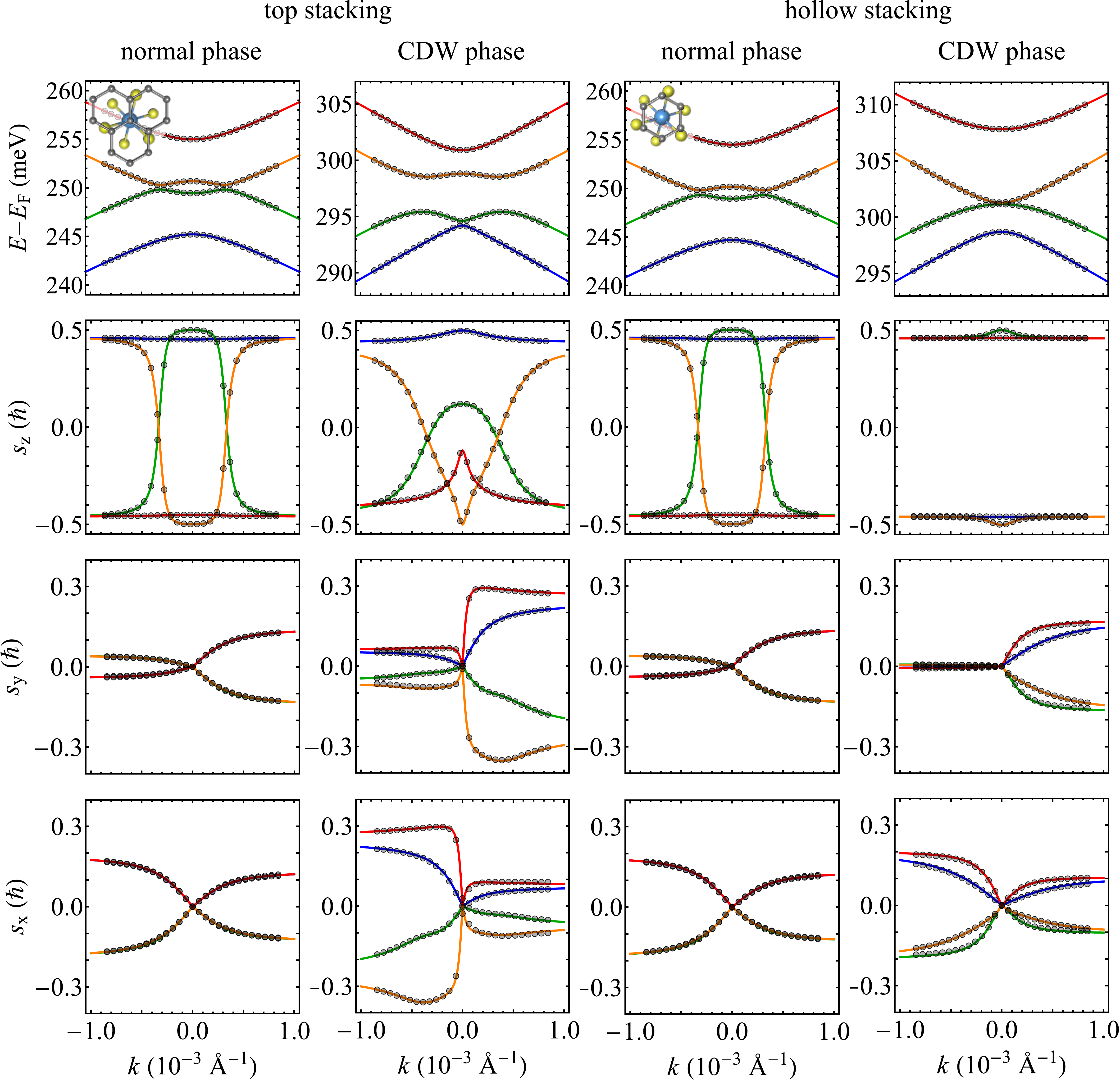}
    \caption{Calculated electronic band structures of graphene on 1T-TaS$_2$ as well as $s_z$, $s_y$, and $s_x$ spin expectation values close to the Dirac point in normal and CDW phase for top and hollow stacking. The solid lines are a tight-binding model fit to the DFT data shown by circles. Each of four line colors corresponds to a distinct band throughout the figure. The wavevectors point towards $\Gamma$ and K, as $\Gamma\leftarrow$~K~$\rightarrow$~M; wavevector $k=0$ corresponds to the K point.} 
    \label{fig3}
\end{figure}

In Figure~\ref{fig2}, we show calculated band structures of the supercell heterostructure for the top stacking along the high symmetry lines in the first Brillouin zone for the case where 1T-TaS$_2$ is in the normal and in the CDW state (including the magnetic state, which we discuss in Section~\ref{Magnetism}).
The fat bands shown by the dark gray circles correspond to the states originating from the carbon atoms. The graphene gets hole-doped and its Dirac cone is shifted about 0.25~eV and 0.30~eV above the Fermi level for the normal and CDW phase, respectively. 
The $\pi$-bands of graphene hybridize with the 1T-TaS$_2$ states in the vicinity of the Fermi level. We note that in the case of the CDW phase, the bands of 1T-TaS$_2$ significantly change and the Dirac states hybridize near the Fermi level with the partially filled Mott band originating predominantly from the $5d$ orbitals of Ta atoms. The change in hybridization has a direct effect on proximity-induced spin-orbit coupling in graphene. 
The band structure for the hollow stacking (not shown) possesses on the energy scale as shown in Fig.~\ref{fig2} almost the same bands' dispersion as the top stacked configuration. The bands differ noticeably in the vicinity of the Dirac point for two stackings in the CDW phase as will be seen from discussion of Fig.~\ref{fig3}.

The Figure~\ref{fig3} shows details of the band structure topology of graphene in the vicinity of the Dirac point and corresponding $s_z$, $s_y$, and $s_x$ spin expectation values owing to proximity effects for the normal and CDW phase of 1T-TaS$_2$. The calculated DFT results near the K point ($k=0$ corresponds to the K point) are shown by circles while the solid lines are tight-binding model fits. Four different colours of these lines mark distinct bands in all the panels of the figure. In each case, both the electron and hole bands are spin split. For the normal phase at the top and hollow stacking, the band dispersions are electron-hole-like symmetric with significant band inversion and avoided crossings due to proximity-induced Rashba spin-orbit coupling. A similar band topology has been observed in the case of graphene on 1H-WSe$_2$ leading to quantum spin Hall effect and pseudohelical states predicted in graphene nanoribbons~\cite{Gmitra2016,GmitraProceedingsPoland}. We note that in the normal phase, the $s_x$ and $s_y$ spin expectation values are identical for pairs of particle-hole symmetric bands, so that they overlap in the plots. 
Also, $s_x$ and $s_y$ spin expectation values vanish exactly at the K point for all cases considered in Fig.~\ref{fig3} due to the inherent $C_3$ symmetry. The emergence of CDW in the underlying 1T-TaS$_2$ monolayer leads to the strong band topology modification due to a change in the spin-orbit coupling proximity-induced effects. 

\begin{figure*}
    \centering
    \includegraphics[width=0.80\columnwidth]{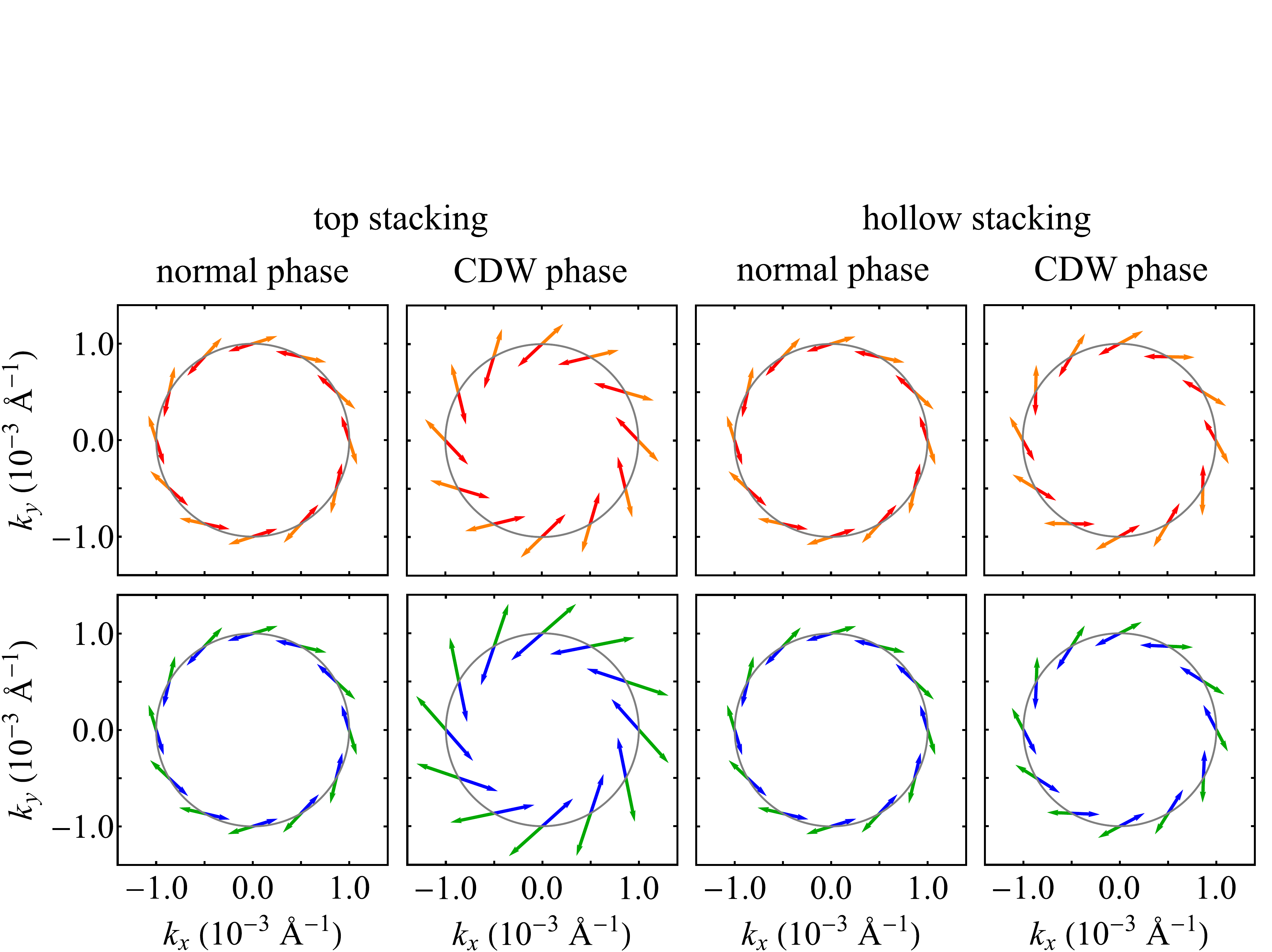}
    \caption{In-plane spin expectation values along the circular contour in wavevector space, centered at the Dirac point K in normal and CDW phase for top and hollow stacking, calculated with DFT. The four arrow colors correspond to the four individual bands as in Fig.~\ref{fig3} (the top row presents data for two conduction-like bands and bottom one for two valence-like bands).}
    \label{fig4}
\end{figure*}

The complex features of the proximity-induced effects on graphene electronic structure can be studied by analyzing an effective tight-binding model. We used the Hamiltonian given by Eq.~(\ref{Eq:Ham}) and fit its parameters to the DFT data.
The parameter values are summarized in Table~\ref{tab:parameters} and provide a very good reproduction of the electronic band structure close to the Dirac point as well as the corresponding spin expectation values; compare DFT data (circles) and the tight-binding model (solid lines) in Fig.~\ref{fig3}.
The effect of the CDW phase in the case of top stacking leads to the band topologies mostly controlled by the sign change of the staggered potential $\Delta$ and significant spin expectation values variation. In the case of hollow stacking with CDW, the bands' dispersions are similar to the case of bare graphene in a strong transverse electric field which induces $|\lambda_{\rm R}|\gg|\lambda_{\rm I}|$~\cite{Gmitra2009} and enforces the degeneracy of a pair of bands at the K point. However, in this case, the bands are nearly degenerate and the physical mechanism is more complex due to the presence of significant staggered potential and the sublattice-asymmetry of the intrinsic spin-orbit coupling (see the parameter values in Table~\ref{tab:parameters}). The most striking effect of the CDW phase is the enhancement of the Rashba angle $\phi_{\rm R}$ in both stackings. This notably cants the in-plane spin components towards the radial direction, see Fig.~\ref{fig4} showing the behavior of the in-plane spin components along the circular contour in wavevector space at the distance of $10^{-3}$~${\rm \AA}^{-1}$ from the K point. It is noteworthy that each arrow color corresponds to an individual energy band as in Fig.~\ref{fig3}. A somehow similar effect regarding the presence of a radial component of the Rashba field was predicted in twisted graphene/TMDC structures~\cite{Li2019b,Naimer2021,Peterfalvi2022}.
In the context of the enhancement of the Rashba angle by the emergence of CDW, let us emphasize the fundamental change in the band structure in the vicinity of Dirac point focusing on the bands originating from TaS$_2$ (compare Fig.~\ref{fig2}(a) and (b)). In the spirit of the model presented in Ref.~\cite{David2019} (discussing the case of a twisted heterostructure of graphene with TMDC), the change of the Rashba angle is a consequence of a pronounced difference between the TMDC energy band structure in the normal phase and the CDW phase.

\begin{table}
\caption{\label{tab:parameters}Tight-binding model parameters for graphene on 1T-TaS$_2$ determined by fitting the model given by Hamiltonian in Eq.~(\ref{Eq:Ham}) to DFT data.} 
\begin{indented}
\lineup
\item[]\begin{tabular}{@{}*{7}{l}}
\br                              
&\centre{2}{top stacking}&\centre{2}{hollow stacking}\\
\ns
&\crule{2}&\crule{2}\\
& normal & CDW & normal & CDW\\
\mr
       $E_0$ (meV)                          & 250.08 & 297.13 & 249.57 & 302.24 \cr
       $t$ (eV)                             & \0\02.547 &\0\02.516 & \0\02.542 &\0\02.537 \cr
       $\Delta$ (meV)                       & \0\01.90 &\0\0\-1.30 &\0\01.91 & \0\02.13 \cr
       $\lambda_{\mathrm {I}}$ (meV)        & \0\0\-0.02 &\0\00.62 &\0\0\-0.02 &\0\0\-1.02 \cr
       $\delta\lambda_{\mathrm {I}}$ (meV)  & \0\0\-2.52 &\0\0\-1.85 &\0\0\-2.52 & \0\0\-2.06 \cr
       $\lambda_{\mathrm R}$ (meV)          & \0\01.05 & \0\01.12 &\0\01.05 & \0\00.90 \cr
       $\phi_{\rm R}$ (deg)                 & \017.36 &\043.07 &\017.57 &\028.15 \cr
\br
\end{tabular}
\end{indented}
\end{table}

\section{Magnetism in graphene on 1T-TaS$_2$ monolayer}\label{Magnetism}

The partially filled band at the Fermi level at the CDW phase of the 1T-TaS$_2$ monolayer is suspected to indicate non-vanishing magnetic polarization according to recent first-principles studies \cite{Jiang2021,Pasquier2022}.
In order to study the presence of spontaneous magnetic polarization in graphene/1T-TaS$_2$ monolayer, we have performed spin-polarized DFT+U calculations considering the CDW phase for 1T-TaS$_2$. The stability of the magnetically ordered state can be evaluated by computing the energy difference between the non-magnetic and the magnetic ground state. Our DFT+U calculations yielded values (per supercell) equal to 34.7~meV and 33.1~meV for top and hollow stacking, respectively. On the other hand, analogous calculations performed for 1T-TaS$_2$ monolayers with the graphene layer removed resulted in values of 29.8~meV and 28.2~meV, respectively. The results support the view that the creation of a heterostructure increases slightly the stability of the magnetic state in 1T-TaS$_2$. For the calculations we have considered a preferential in-plane magnetization along the $x$-direction, as we found that the  magnetic anisotropy energy, i.e., the in-plane minus out-of-plane configuration energy (per supercell) amounts to 0.56 meV and 0.76 meV for top and hollow stacking, respectively, supporting the easy-plane anisotropy. The same type of anisotropy is predicted to emerge for a bare 1T-TaS$_2$ monolayer with magnetic moments centered on David stars \cite{Pasquier2022}. 

The band structure resulting from DFT+U calculations with CDW phase in 1T-TaS$_2$ is shown in Fig.~\ref{fig2}(c), where the fat bands originating from Ta atoms are marked with circles with color scale representing the degree of spin-polarization, linear color scale from red color for spin-up polarization, through gray to blue color for spin-down polarization. The main feature distinguishing this case from unpolarized CDW phase shown in Fig.~\ref{fig2}(b) is pronounced spin splitting of the Hubbard flat Ta band located close to the Fermi level to a pair of strongly polarized bands. The remaining bands for the plotted energy range exhibit significantly lower degree of polarization.

\begin{table}
\caption{\label{tab:CDW_ex}
Tight-binding model parameters for graphene on 1T-TaS$_2$ in CDW phase and in-plane magnetization determined by fitting the model given by Eq.~(\ref{Eq:Ham}) and (\ref{Eq:Ham_ex}) to DFT data.
} 
\begin{indented}
\lineup
\item[]\begin{tabular}{@{}*{3}{l}}
\br                              
&top stacking&hollow stacking\cr 
\mr
$E_0$ (meV)                         &336.50 &336.16\cr
$t$ (eV)                            &\0\02.516&\0\02.537\cr
$\Delta$ (meV)                      &\0\01.81 &\0\0\-2.27\cr
$\lambda_{\mathrm {I}}$ (meV)       &\0\01.22 &\0\0\-1.83\cr
$\delta\lambda_{\mathrm {I}}$ (meV) &\0\01.98 &\0\02.47\cr
$\lambda_{\mathrm R}$ (meV)         &\0\01.65 &\0\01.07\cr
$\phi_{\rm R}$ (deg)                    &\056.90 &\030.93\cr
$\Delta_{\rm A}$ (meV)                  &\0\0\-2.33 &\0\0\-0.85\cr
$\Delta_{\rm B}$ (meV)                  &\0\0\-1.66 &\0\0\-0.74\cr
$\phi_{\rm A}$ (deg)                    &\0\0\-9.31 &\0\0\-5.90\cr
$\phi_{\rm B}$ (deg)                    &\0\0\-3.86 &\0\-20.37 \cr
\br
\end{tabular}
\end{indented}
\end{table}

\begin{figure*}
    \centering
    \includegraphics[width=0.6\columnwidth]{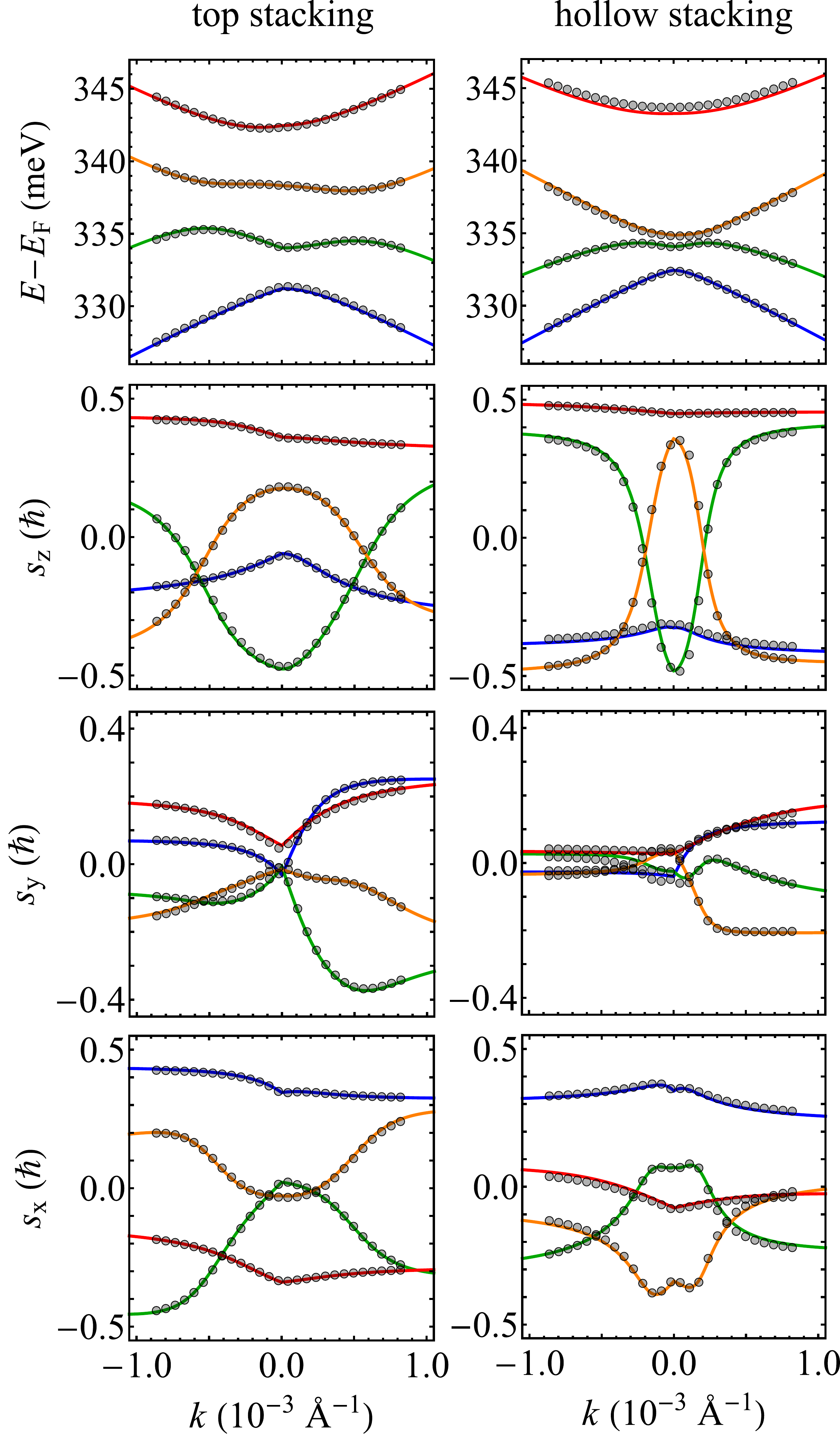}
    \caption{Electronic band structures and spin expectation values of $s_z$, $s_y$ and $s_x$ close to the Dirac point in CDW phase for top and hollow stacking, calculated with DFT+U. The solid lines are tight-binding model fit to the DFT+U data shown by solid circles. Each of four line colors corresponds to a distinct band throughout the figure. The $k$ vectors are centered around the K point, $k=0$, heading towards $\Gamma\leftarrow$~K~$\rightarrow$~M.} 
    \label{fig5}
\end{figure*}

\begin{figure*}
    \centering
    \includegraphics[width=0.6\columnwidth]{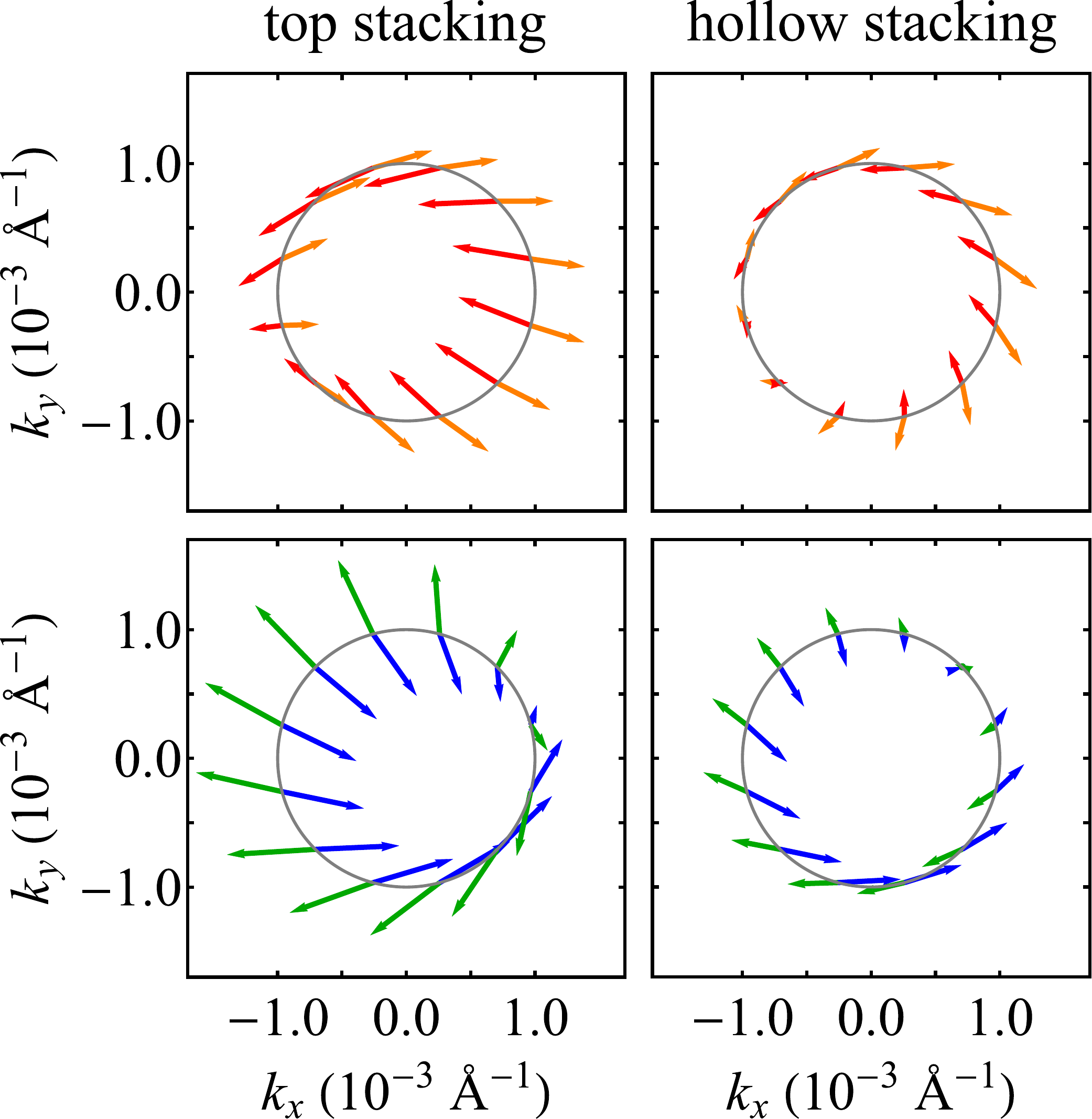}
    \caption{In-plane spin expectation values along the circular contour in wavevector space, centered at the Dirac point K for top and hollow stacking, calculated with DFT+U. The four arrow colors correspond to the four individual bands as in Fig.~\ref{fig5} (top row presents data for two conduction-like bands and bottom one for two valence-like bands).}
    \label{fig6}
\end{figure*}

The effect of the in-plane magnetization on the graphene bands in the vicinity of the K/K' point is well described by the effective Zeeman-like Hamiltonian $H_{\rm ex}$, given by Eq.~\eqref{Eq:Ham_ex}, representing the coupling of a spin with the sublattice-dependent in-plane magnetization, having the interaction strength 
$\Delta_{\rm A/B}$ and an angle $\phi_{\rm A/B}$
with respect to the considered $x$-direction. Using the Hamiltonian $H_\kappa+H_{\rm ex}$ we were able to fit relativistic non-collinear DFT data. The obtained parameters are given in Table~\ref{tab:CDW_ex}, while the comparison between the DFT and fitted data in Fig.~\ref{fig5} shows that the addition of the $H_{\rm ex}$ is able to capture the dominant behavior of the graphene bands in the vicinity of the K/K' point when the magnetic polarization emerges. In both the top and the hollow stacking, the sublattice magnetization direction $\phi_{\rm A/B}$ mostly points towards the $x$-direction. It resembles the situation found in monolayer 1T-TaS$_2$, in which the Mott-like band that lies on the Fermi level strongly contributes to the $x$-component of magnetization \cite{Pasquier2022}.
The in-plane spin pattern, shown in Fig.~\ref{fig6}, differs significantly from the one depicted in Fig.~\ref{fig4}, for the non-magnetic case with CDW. The tilt angle shows pronounced angular dependence and, also, differs significantly for all four bands involved.

Our fitting shows that the presence of magnetism influences the graphene bands not only qualitatively through the appearance of the effective Zeeman-like Hamiltonian, but also quantitatively through the change of the nonmagnetic parameters of graphene Hamiltonian (Eq.~\eqref{Eq:Ham}). This is not a surprise, since the exchange parameters $\Delta_{\rm A/B}$ of the order of 1~meV correspond to the giant built-in magnetic field of 17 T (see a comparison between  the nonmagnetic data in Table~\ref{tab:parameters} and magnetic data in Table~\ref{tab:CDW_ex}),
thus justifying the observed interference between the proximity-induced spin-orbit coupling and proximity-induced-magnetism. The effective model Hamiltonian we developed can be used in future studies probing interplays of several proximity-induced interactions, for instance, investigating spin-to-charge inter-conversion phenomena~\cite{Sousa2020} or novel anisotropic Rashba-Edelstein effects that can be tuned by a twist \cite{Veneri2022} or by the engineering of CDW. The temperature in such cases can be used as an easy knob for switching between different phases and for turning on/off spin-orbit and exchange interactions.

\section{Final remarks}

In the paper, we report a study of the electronic properties of van der Waals graphene/1T-TaS$_2$ heterostructure focusing on the proximity effects in the vicinity of the graphene Dirac point. The underlying band structures were computed by DFT calculations and then modeled analytically employing an effective tight-binding Hamiltonian derived from the symmetry considerations. The model parameters were determined by fitting the tight-binding to the DFT data. A crucial finding is the fundamental influence of CDW formation on the proximity effect, resulting in a profound change in the parameters, spin textures, and fine-band topologies. As a consequence, controlling the CDW presence in a heterostructure consisting of graphene on 1T-TaS$_2$ provides a promising tool to manipulate the electronic properties of graphene near the Dirac point, in particular, a possibility to tailor the radial Rashba SOC, which cants the in-plane spin projections. In addition to SOC proximity, the in-plane magnetic polarization of the CDW phase superimposes the SOC, and exchange proximity effects offering to control the proximity effects by temperature.

Our findings demonstrate that the emergence of CDW in 1T-TaS$_2$ profoundly modifies the proximity effects in the studied heterostructure with graphene. The effect bears resemblance to the physical situation in twisted heterostructures, where modification of the twist angle allows control over the proximity effects. It should be strongly emphasized that in our system this effect takes place without physical modification of the relative twist angle of 13.9$^{\circ}$ between the graphene layer and 1T-TaS$_2$ layer, as enforced by the lattice commensurability for $\sqrt{13}\times\sqrt{13}$ reconstruction. Therefore, a specific sort of  spintronics/twistronics without a twist may be realized based on CDW control in our heterostructure. 

The proximity-induced SOC controlled with the twist angle in van der Waals graphene/TMDC heterostructure might find application in spin-to-charge inter-conversion~\cite{Ghiasi2019,Veneri2022}. In particular, the modification of the in-plane spin texture with the twist angle enables tunability of the Rashba-Edelstein effect \cite{Veneri2022}. In this context, we might mention that the controllability of the proximity effects in our heterostructure offers a route to master also the spin-to-charge inter-conversion but without the need to tune physically the twist angle and with full reversibility.

\ack
\textbf{K.Sz.}~acknowledges financial support provided by the National Science Centre (Poland) under Grant No.~2015/19/B/ST3/03142 and by University of Lodz under Grant No.~1/IDUB/DOS/2021.
\textbf{M.M.}~acknowledges financial support provided by the Ministry of Education, Science, and Technological Development of the Republic of Serbia and DAAD Research Grant 57552336. Also, this project has received funding from the European Union's Horizon 2020 Research and Innovation Programme under the Programme SASPRO 2 COFUND Marie Sklodowska-Curie grant agreement No.~945478.
\textbf{D.K.}~acknowledges partial support from the project SUPERSPIN funded by Slovak Academy of Sciences within the program IMPULZ IM-2021-26, as well as
from Deutsche Forschungsge\-meinschaft (DFG, German Research Foundation) within Project-ID 314695032-SFB 1277, and the Grant No.~2/0092/21 awarded by the Ministry of Education, Science, Research and Sport of the Slovak Republic. 
\textbf{M.G.}~acknowledges financial support from Slovak Research and Development Agency provided under Contract No. APVV SK-CZ-RD-21-0114 and APVV 20-0425, and by the Ministry of Education, Science, Research and Sport of the Slovak Republic provided under Grant No. VEGA 1/0105/20 and Slovak Academy of Sciences project IMPULZ IM-2021-42 and project FLAG ERA JTC 2021 2DSOTECH.

\section*{ORCID IDs:}

Karol Sza\l{}owski: https://orcid.org/0000-0002-3204-1849\\
Marko Milivojevi{\'c}: https://orcid.org/0000-0002-9583-3640\\
Denis Kochan: https://orcid.org/0000-0003-0613-5996\\
Martin Gmitra: https://orcid.org/0000-0003-1118-3028



\end{document}